\documentclass[journal]{IEEEtran}
\usepackage[pdftex]{graphicx}

\begin{document}
%
\title{Comparison of 3 absolute gravimeters based on different methods for the e-MASS project}

\author{
Anne~Louchet-Chauvet, S\'ebastien~Merlet, Quentin~Bodart, Arnaud~Landragin and Franck~Pereira~Dos~Santos,\\
Henri Baumann,\\
Giancarlo D'Agostino and Claudio Origlia

\thanks{A. Louchet-Chauvet, S. Merlet, Q. Bodart, A. Landragin and F. Pereira Dos Santos are with LNE-SYRTE, CNRS UMR 8630, UPMC, 61 avenue de l'Observatoire, 75014 Paris, France.}
\thanks{H. Baumann is with Federal Office of Metrology METAS, Lindenweg 50, CH-3003 Wabern-Bern, Switzerland.}
\thanks{G. D'Agostino and C. Origlia are with Istituto Nazionale di Ricerca Metrologica INRIM (formerly IMGC-CNR), Strada delle Cacce 73, 10135 Turin, Italy.}
}


\maketitle

\begin{abstract}
We report on the comparison between three absolute gravimeters
that took place in April 2010 at Laboratoire National de
M\'etrologie et d'Essais. The three instruments (FG5\#209 from
METAS, Switzerland, IMGC-02 from INRIM, Italy, and CAG from
LNE-SYRTE, France) rely on different methods: optical and atomic
interferometry. We discuss their differences as well as their
similarities. We compare their measurements of the gravitational
acceleration in 4 points of the same pillar, in the perspective of
an absolute determination of $g$ for a watt balance experiment.
The three instruments performed repeatable $g$ measurements, but
do not agree at the level aimed for. This work calls for
additional studies on systematic effects.
\end{abstract}

%
\IEEEpeerreviewmaketitle

\section{Introduction}
The National Metrology Institutes of Switzerland (METAS), Italy
(INRIM) and France (LNE) are involved in the e-MASS Euramet Joint
Research Project~\cite{geneves:emass}. This project aims at giving
a new definition of the kilogram with the help of a watt balance,
which weighs a reference mass in terms of electrical
quantities~\cite{kibble}. In the gravimetry section of this
project, two tasks have been identified: (i) determine the value
of the gravitational acceleration with absolute gravimeters, and
(ii) transfer the absolute value to the position of the test
mass~\cite{merlet2008,baumann2009}. The objective we pursue is to
reach an accuracy at the $\mu$Gal level on both tasks. This paper
reports on our efforts towards the completion of the first one.
Although conventional optical gravimeters allow routine
measurements of $g$ with repeatabilities of the order of a few
$\mu$Gal~\cite{merlet2007}, comparisons with instruments based on
other technologies are desirable to confirm the accuracy of their
measurements, especially in the context of the possible
redefinition of SI units.

To that end, the three institutes operate three different absolute
gravimeters which rely on different measurement methods. METAS
operates a commercial absolute gravimeter FG5\#209 from Micro-g
Lacoste, relying on the direct free-fall method. INRIM has been
developing its own ballistic gravimeter IMGC-02 based on the
symmetrical rise-and-fall method. LNE-SYRTE has been developing a
cold atom gravimeter (CAG) based on atom interferometry, to be
used with the watt balance in construction at Laboratoire National
de M\'etrologie et d'Essais (LNE)~\cite{geneves:bw}. We report on
a comparison involving these three devices, organized at LNE in a
room dedicated to gravimetry, next to the watt balance room.

\section{Presentation of the three gravimeters}
The three gravimeters involved in the comparison are national
references for their respective country. All of them are based on
tracking the trajectory of a free-falling test mass with a laser,
using an interferometric method. However, they use different
methods to measure $g$. Although they all involve
vibration-rejection systems, their sensitivities are still limited
by mechanical vibrations.

\subsection{Optical gravimeter FG5}
The FG5\#209 absolute gravimeter of METAS is a state-of-the-art commercial gravimeter~\cite{niebauer:fg5}. It is essentially a modified Mach-Zehnder interferometer, in which one arm is reflected on a free-falling corner cube.
The corner cube is placed in a carriage that is lifted to the top of an evacuated dropping chamber. The carriage then accelerates downwards with an acceleration higher than $g$, to allow a $20$~cm free fall of the corner cube. The trajectory is sampled by counting the interference fringes at the output of the interferometer. The laser source used is a HeNe laser frequency-stabilized on an iodine reference. The corner cube is then lifted again to the top of the chamber to prepare for the next measurement.
To reduce the influence of ground vibrations, a reference corner cube is fixed to an active inertial reference (Super Spring)~\cite{nelson:superspring}.

FG5 devices are used to determine the free fall acceleration in all the watt balance experiments~\cite{eichenberger2009} except at LNE~\cite{geneves:bw}.

\subsection{Optical gravimeter IMGC-02}

INRIM has been developing its own absolute optical gravimeter. The IMGC-02 is also an optical interferometer involving a iodine-stabilized HeNe laser. Unlike the FG5, it uses the symmetric rise-and-fall method: the test mass, a corner cube, is thrown vertically upwards in an evacuated chamber. A reference corner cube is fixed to the inertial mass of a long-period seismometer. The trajectories are reconstructed by sampling the fringes with a digital oscilloscope~\cite{dagostino2005}. The acceleration experienced by the falling corner cube is determined by fitting a motion model to the tracked trajectory~\cite{dagostino2008}.

\subsection{Atomic gravimeter CAG}
The cold atom gravimeter developed at LNE-SYRTE uses atom interferometry to perform a cyclic absolute measurement of $g$. At each cycle, a new cold cloud of Rb atoms is prepared in a UHV chamber, to be used as a test mass.
During their free fall, these atoms undergo three stimulated Raman transitions that respectively separate, redirect, and recombine the atomic wave function, resulting in an atomic interferometer. The total phase shift between the two paths of this atomic interferometer depends on $g$, and scales with the square of the time interval between two consecutive Raman pulses. This gravity phase shift is cancelled by chirping the frequency difference of the two Raman beams in order to compensate the time-dependent Doppler shift. The value of $g$ is therefore derived from a frequency chirp.

\begin{figure}[!t]
\centering
\includegraphics[width=3.0in,bb=19 117 579 725]{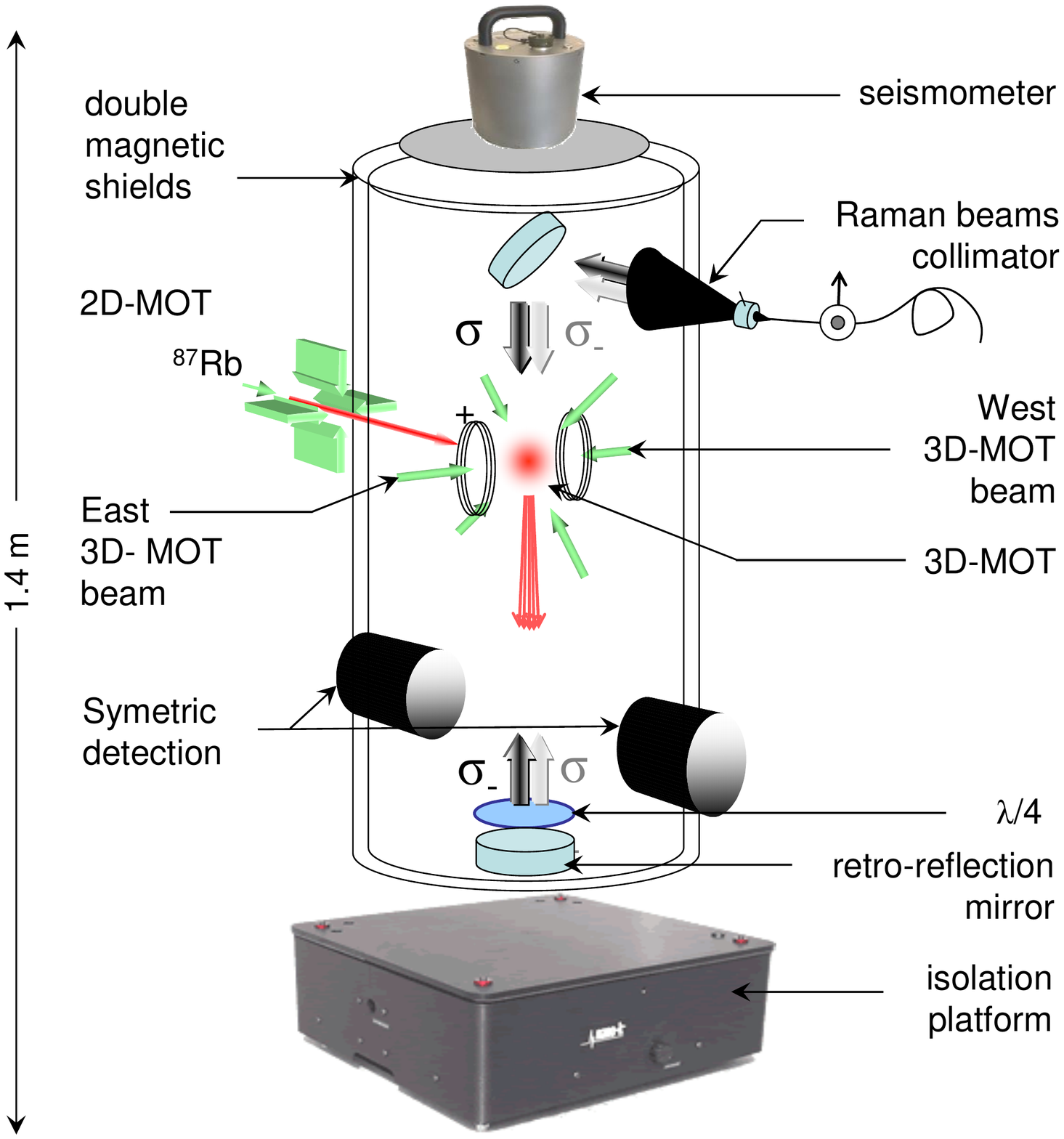}
\caption{Schematic of the cold atom gravimeter (CAG). A cold
$^{87}$Rb atomic sample is prepared by 2D- and 3D-magneto-optical
trapping. After a short sub-Doppler cooling phase, where the atoms
reach a temperature of $\sim2\mu$K, the trapping beams are
switched off and the atoms start their free fall in the vacuum
chamber. During the free fall, a sequence of three stimulated
Raman transitions is used to realize the atomic interferometer.
These transitions are performed with two vertical,
counter-propagating, circularly polarized laser beams, addressing
the hyperfine transition of rubidium at $6.834$~GHz via two-photon
excitation. The phase shift at the output of the interferometer is
deduced from a symmetric fluorescence measurement of the atomic
state at the bottom of the vacuum chamber~\cite{bordé1989}.}
\label{fig:cag}
\end{figure}

A more complete description of the CAG is given in
Figure~\ref{fig:cag}. The vacuum chamber lies on top of a passive
isolation platform. The non-filtered vibration noise is measured
with a Guralp seismometer rigidly attached to the vacuum chamber,
and is used to post-correct the atomic signal~\cite{legouet2008}.

The device used in this comparison is an improved version of the
prototype gravimeter described in~\cite{legouet2008}. The vacuum
chamber is now made of titanium, in order to minimize magnetic
field gradients and Eddy currents. The retroreflecting mirror for
the Raman beams is placed inside the chamber, leading to reduced
optical wavefront aberrations. Furthermore, the fluorescence
detection is performed with a double set of detectors placed
symmetrically at opposite sides of the atomic cloud.

Although the CAG is by far the largest of the 3 gravimeters, it is
nevertheless transportable and can be moved from one point to
another in a room in about 2 hours.

\section{Details about the comparison}

The comparison was carried out between April 11th and April 21st
2010 in the Laboratoire National de M\'etrologie et d'Essais
(LNE). The three devices measured gravity in different points of
the GR room next to the BW room, where the watt balance is being
developed at LNE.

\begin{figure}[!t]
\centering
\includegraphics[width=3.2in]{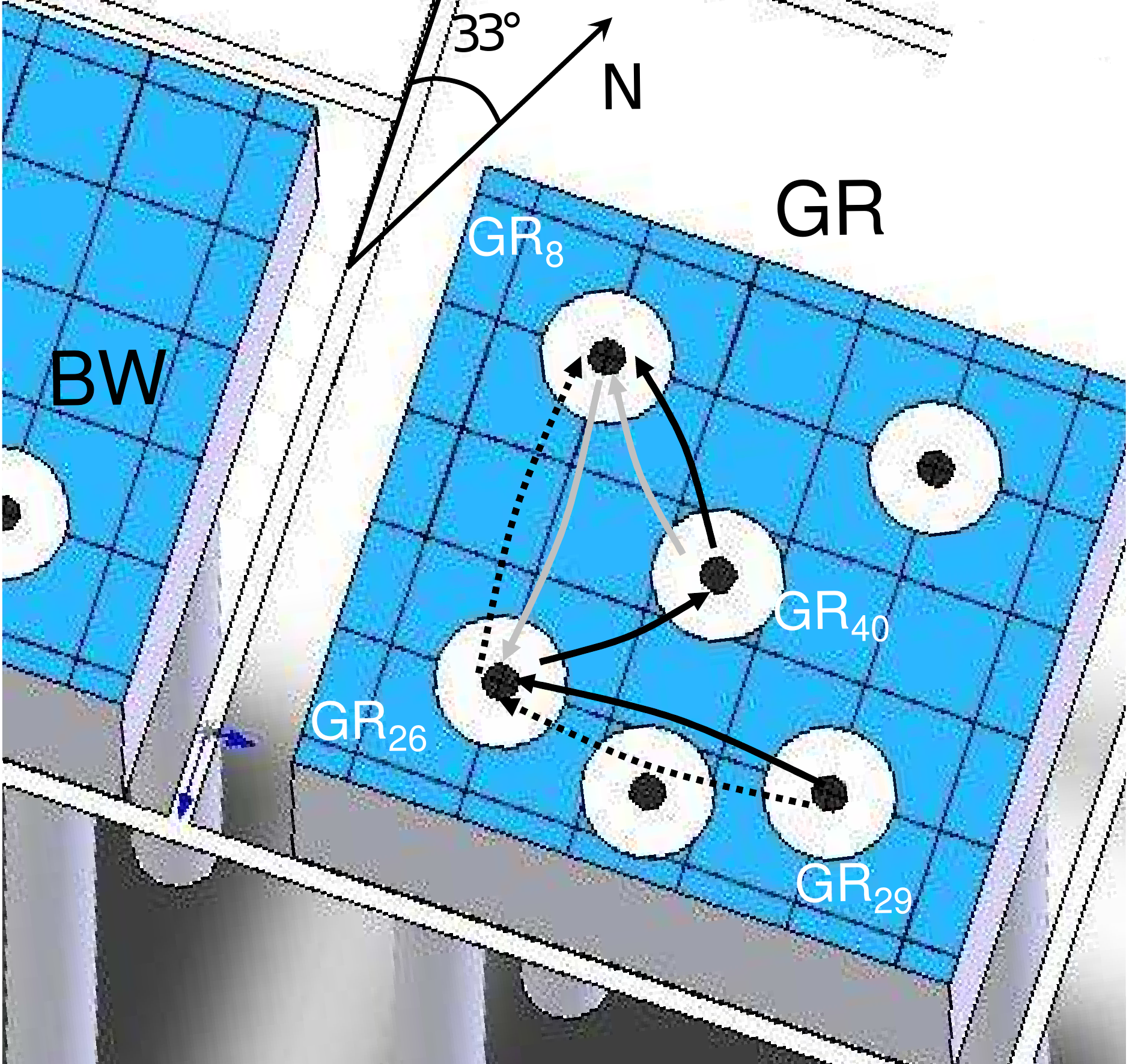}
\caption{Sequence of the comparison in the GR room at LNE, from
April 11th 2010 (MJD=55297) to April 21st 2010 (MJD=55307). The
different devices' sequences are symbolized with dashed, gray, and
black arrows, for FG5\#209, CAG and IMGC-02, respectively. The LNE
watt balance is located in the next room (BW), on a similar
pillar.} \label{fig:seq}
\end{figure}

Figure~\ref{fig:seq} illustrates the schedule of the comparison
procedure, as well as the room configuration. Gravity was measured
in four different locations on the pillar, denoted GR40, GR8, GR26
and GR29. Each gravimeter measured $g$ in at least three out of
these four points. Extensive relative gravimetry characterizations
had previously been performed in the GR room~\cite{merlet2008}. In
particular we had measured the ties between the different points
as well as the vertical gravity gradients with the commercial
$Scintrex$ CG5-S105 relative gravimeter. We measured again the
vertical gravity gradients right after the comparison at the four
measurement points, following the same procedure, and found
results in perfect agreement with the previous determination.

The $g$ values measured with the METAS FG5\#209 are given at a
height of $122$~cm above ground. Every $30$ minutes, the device
performs a series of 100 drops spaced by $10$~s. The IMGC-02
gravimeter performs $g$ measurements $47.2~$cm above ground, at a
rate of one throw every $30~$s, only during nighttime where the
environmental noise is weaker. Concerning the CAG, a measurement
of $g$ is achieved every $0.36$~s, which corresponds to a
repetition rate of $2.8$~Hz. The measurement height is $83.5$~cm
above ground.

For the final comparison results, the $g$ values will be given at
the height of $84.25~$cm above ground, which is the mean height of
the 3 devices. The transfer of $g$ is calculated from the measured
vertical gravity gradients.

\section{Results}

\subsection{Gravity measurements}
\begin{figure*}[!t]
\centering
\includegraphics[width=7.2in,angle=0, bb=0 30 841 478]{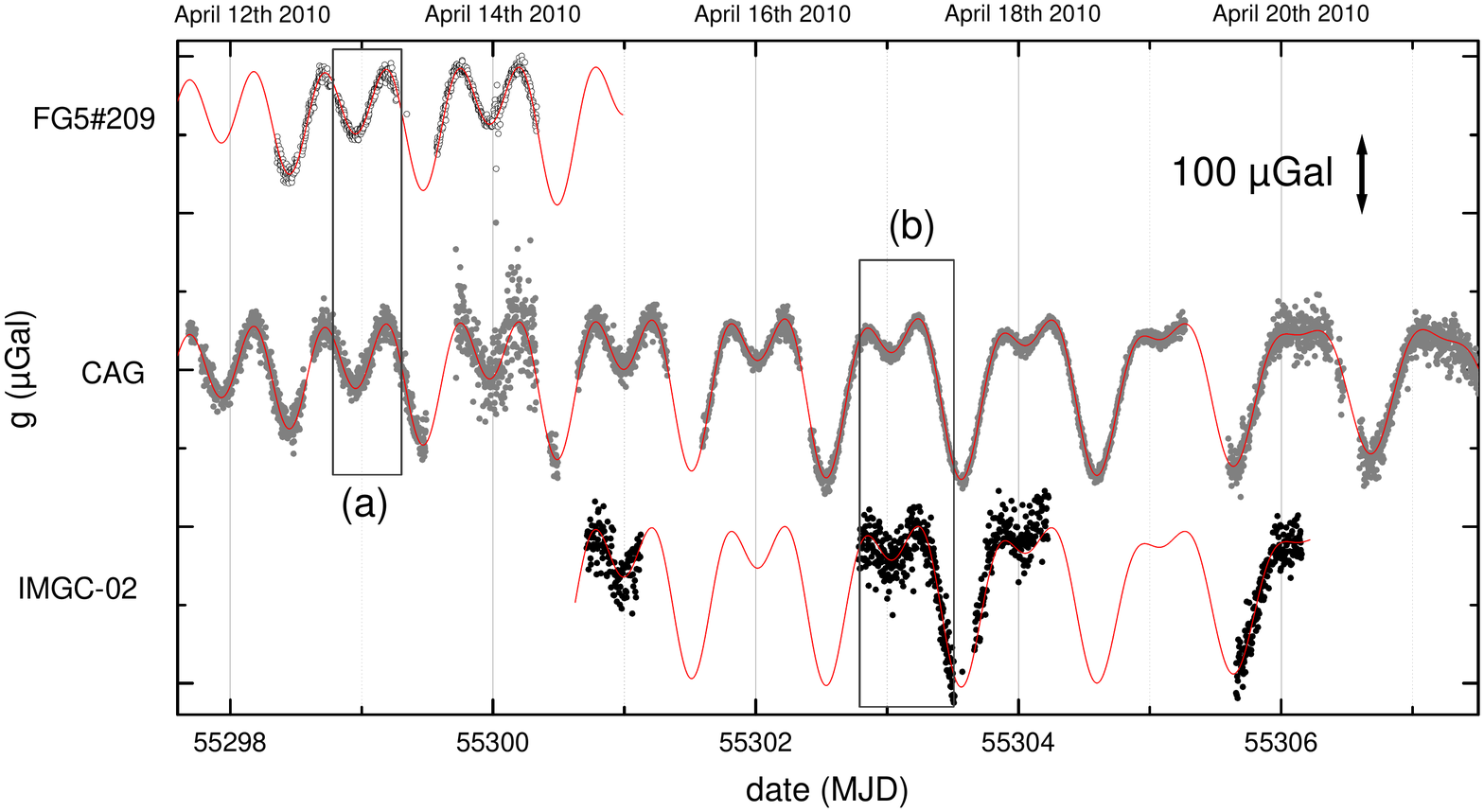}
\caption{$g$-measurements over the duration of the comparison,
from April 11th to April 21st 2010. Open circles: FG5\#209 (bins
of 8 drops); Gray circles: CAG (bins of 400 drops); Black circles:
IMGC-02 (bins of 5 throws); Solid red line: tide model. The $g$
values are vertically offset between different devices for
clarity. The rectangles denoted (a) and (b) indicate the time
period where the Allan deviations represented in
Figure~\ref{fig:allan} have been computed.} \label{fig:goft}
\end{figure*}

Figure~\ref{fig:goft} displays the non-corrected $g$ measurements
over the whole duration of the comparison, with a binning time of
$150~$s corresponding respectively to $8$ drops for FG5, $5$
throws for IMGC-02, and $400$ drops for CAG.

For the FG5\#209 measurements, errors in the fringe counting
process led to rejecting $18$\% of the data points lying far off
from the center of the statistical distribution, otherwise fairly
gaussian.

For the IMGC-02, the outliers rejection method is based on the
Chauvenet criterion. The main part of the rejected data
corresponds to excessive variations of the fringe visibility
during the launch. This amounts to a rejection of $47$\% of the
data points.

The cold atom gravimeter performed almost continuous $g$
measurements throughout the 10 days of the comparison. The only
interruptions in the data acquisition correspond either to optical
realignment and laser relocking sessions, or to the transportation
of the device to another point in the room. A $3\sigma$ rejection
scheme would discard about $0.6$\% of the data points, i.e. of the
order of what one would obtain with a normal distribution. No
rejection scheme was finally applied for the CAG measurements.

The noise in the CAG measurements appears to be changing over the
course of the comparison. Although we cannot rule out the
possibility that the vibration noise depends on the position of
the instrument in the room, it is more likely that the noise level
variations are due to more or less efficient vibration isolation.
The particularly noisy measurements performed on
$\textrm{MJD}=55300$ can be explained by an imperfect adjustment
of the isolation platform, as well as non-optimal vibration
correction parameters. Conversely, the quiet measurements
performed from $\textrm{MJD}=55303$ to $55305$ can be attributed
to a combination of favorable conditions: good weather, little
human activity (saturday and sunday), excellent correction of
vibrations, and also relatively small power fluctuations in the
Raman beams.\label{noises}

\subsection{Corrections to the $g$ measurements }
To get an absolute measurement of $g$, we correct the measurements
from the usual environmental perturbations: polar motion,
atmospheric pressure, tides and ocean loading, using different
models that agree with each other. The three instruments also
suffer from systematic effects, such as alignment, optical beam
quality, self gravity, Coriolis effect, or reference frequency
offset, for instance.

The instrument-specific corrections and corresponding
uncertainties for the optical gravimeters FG5 and IMGC-02 are
taken from references~\cite{niebauer:fg5} and~\cite{dagostinoPhD},
respectively. For the CAG, we give a more complete description of
the corrections that are the most delicate to evaluate, namely the
Coriolis effect, optical  aberrations in the Raman beams, and
two-photon light shift.

The non-zero initial velocity of the atomic cloud in the East-West
direction gives rise to a bias on the gravity measurement coming
from the Coriolis force. To estimate this bias we rotate the cold
atom gravimeter by $180^\circ$ around the vertical axis. In these
two configurations the contribution of the Coriolis effect has the
same amplitude but opposite sign. The gravity measurements are
therefore corrected with half of the difference between North and
South configuration: $(-1.5\pm0.5)~\mu$Gal.

A non-plane transverse wavefront of the Raman laser beams induces
a bias on the $g$ measurement~\cite{fils2005}. This wavefront is
not well known, although much closer to a plane than in the
prototype version of the gravimeter~\cite{legouet2008}. The
correction to $g$ due to optical aberrations is estimated by
measuring the dependence of $g$ to atomic temperature, for
temperatures ranging from $2~\mu$K to $10~\mu$K. Indeed, the
higher the temperature, the more the atomic cloud expands
transversally and a larger area of the optical wavefront is
probed. An unbiased $g$ value should be obtained for a
non-expanding atomic cloud, corresponding to zero temperature. We
therefore extrapolate the bias to $0~\mu$K and get a correction of
$(0\pm6)~\mu$Gal. This effect is the main contribution to the
uncertainty budget. A more thorough investigation of the optical
aberrations is necessary, and will require a better control of the
initial position and velocity distribution of the atomic cloud.

The two-photon light shift due to the Raman light pulses displaces
the atomic levels and therefore modifies the hyperfine transition
frequency~\cite{gauguet2008}. Over the whole duration of the
comparison, the bias on $g$ induced by the two-photon light shift
varied between $8.9$ and $15.5~\mu$Gal, with an associated
uncertainty of $0.5~\mu$Gal. In order to optimize long-term
stability, this effect is continuously monitored by using four
interlaced successive measurement
configurations~\cite{gauguet2008}. Thus the corrected $g$ value is
obtained from a linear combination of the four measurements, which
scales down the repetition rate to $0.7$~Hz, and finally
deteriorates the sensitivity of the $g$ measurement by
$\sqrt{10}$.

\subsection{Absolute gravity measurements}

\begin{figure}[!t]
\centering
\includegraphics[width=3.2in]{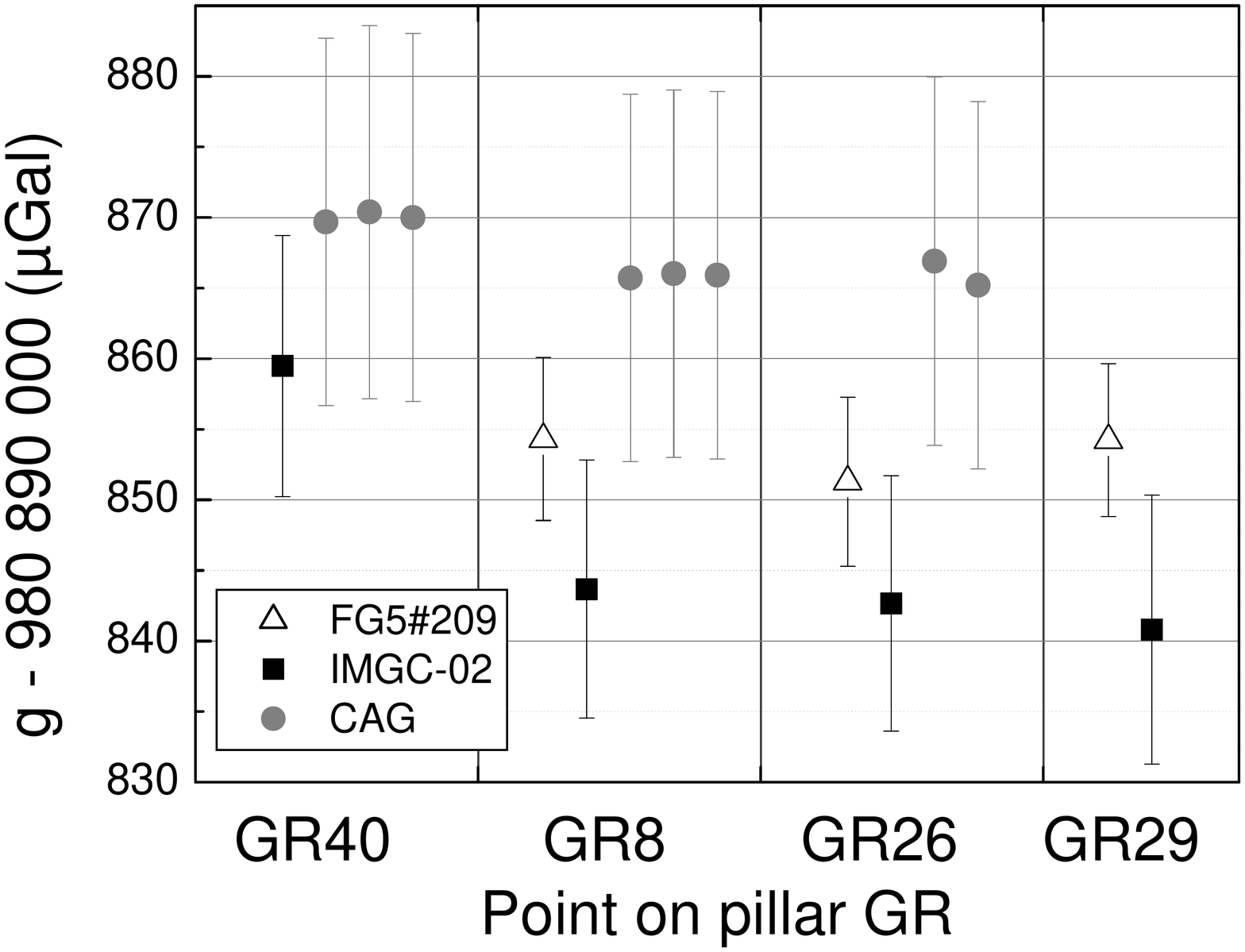}
\caption{Absolute $g$ measurements as given by the three
gravimeters involved in the comparison. The errors bars are given
with $k=2$.} \label{fig:agreement}
\end{figure}

In Table~\ref{tab:g} and Figure~\ref{fig:agreement} we show the
result of the $g$ measurements performed by the 3 absolute
gravimeters, transferred at $84.25~$cm above ground. In the table
we give the experimental standard deviation of the mean value
$s_{gm}$ and the measurement combined uncertainty $u_{gm}$, at
each gravimeter's height of measurement. We also specify
$u_{tie}$, the uncertainty due to the transfer of $g$ to the
height of $84.25$~cm. In Figure~\ref{fig:agreement}, the total
expanded uncertainty is given by $U=k\sqrt{u_{gm}^2+u_{tie}^2}$
with $k=2$. For the 3 instruments, $u_{tie}$ is a negligible
contribution to the total uncertainty.

\begin{table}[!t]
\caption{$g$ values measured 84.25cm above ground expressed in
$\mu$Gal. $u_{tie}$ is the uncertainty due to the transfer of $g$
to the height of $84.25$~cm; $s_{gm}$ is the experimental standard
deviation of the mean value; $u_{gm}$ is the measurement combined
uncertainty.} \label{tab:g} \centering
\begin{tabular}{|c| c| c| c| c| c| c|}
\hline
point & MJD & gravimeter  & $g$ & $u_{tie}$ & $u_{gm}$ & $s_{gm}$ \\
\hline
GR40  & 55298.6 & CAG      & $980\ 890\ 869.7$ & $0.3$ & $6.5$ & $0.4$\\
      & 55300.0 & CAG      & $980\ 890\ 870.4$ & $0.3$ & $6.6$ & $1.2$\\
      & 55301.0 & CAG      & $980\ 890\ 870.0$ & $0.3$ & $6.5$ & $0.5$\\
      & 55304.0 & IMGC-02  & $980\ 890\ 859.5$ & $1.4$ & $4.4$ & $1.5$\\
\hline
GR8  &  55300.0 & FG5\#209 & $980\ 890\ 854.3$ & $1.4$ & $2.5$ & $1.5$\\
     &  55301.9 & CAG      & $980\ 890\ 865.7$ & $0.3$ & $6.5$ & $0.3$\\
     &  55303.0 & CAG      & $980\ 890\ 866.0$ & $0.3$ & $6.5$ & $0.2$\\
     &  55304.4 & CAG      & $980\ 890\ 865.9$ & $0.3$ & $6.5$ & $0.2$\\
     &  55305.9 & IMGC-02  & $980\ 890\ 843.7$ & $1.4$ & $4.3$ & $1.4$\\
\hline
GR26 &  55298.9 & FG5\#209 & $980\ 890\ 851.3$ & $1.4$ & $2.6$ & $1.7$\\
     &  55303.1 & IMGC-02  & $980\ 890\ 842.7$ & $1.4$ & $4.3$ & $1.2$\\
     &  55306.0 & CAG      & $980\ 890\ 866.9$ & $0.3$ & $6.5$ & $0.5$\\
     &  55307.0 & CAG      & $980\ 890\ 865.2$ & $0.3$ & $6.5$ & $0.5$\\
\hline
GR29 &  55298.4 & FG5\#209 & $980\ 890\ 854.2$ & $1.4$ & $2.3$ & $1.1$\\
     &  55300.9 & IMGC-02  & $980\ 890\ 840.8$ & $1.4$ & $4.6$ & $2.0$\\
\hline
\end{tabular}
\end{table}

Without considering the IMGC-02 measurement in GR40, the gravity
variations on the GR pillar measured by each device agree with the
model~\cite{merlet2008} and with the ties determined with the CG5.
For example, $g_\textrm{GR8}-g_\textrm{GR40}=-4.1~\mu$Gal for CAG,
in agreement with the difference of $-4.5~\mu$Gal obtained with
the model. However, for IMGC-02, this difference is as large as
$-17~\mu$Gal, attributed to a laser malfunction observed while
measuring on point GR40.

On the one hand, the instruments allow repeatable $g$
determinations over the duration of the comparison. On the other
hand, absolute measurements obtained with the 3 instruments are
not in full agreement, as shown in Figure~\ref{fig:agreement}. The
difference between absolute measurements can reach as much as
$24.2~\mu$Gal (CAG and IMGC-02, at point GR26), which is larger
than the expanded uncertainty. These differences are related to
systematic effects that remain to be evaluated more carefully,
thanks to more comparisons and studies. Additionally, the results
of the last International Comparison of Absolute Gravimeters
(ICAG'09), where CAG, FG5\#209 and IMGC-02 were present, will
bring complementary information on the repeatability of these
differences.

Gravity on points GR40 and GR29 was determined during two previous
comparisons. In october 2006, $g$ measurements obtained with three
FG5 (\#215, \#216 and \#228)~\cite{merlet2007} showed constant
differences between measurements on same points up to $10~\mu$Gal.
On GR29, the difference between the 2006 mean value and the
FG5\#209 value obtained here is $0.4~\mu$Gal.

The CAG participated in the second comparison with
FG5\#220~\cite{merlet2010} in october 2009. The difference between
FG5\#209 obtained here and FG5\#220 obtained in 2009 is
$1.6~\mu$Gal on GR29. During this last comparison, CAG performed a
$g$ determination on point GR40. The difference between the CAG
measurement obtained here and the one obtained in 2009 is
$15~\mu$Gal at a height of $120$~cm. We believe that this
discrepancy arises from relatively large -but bounded-
fluctuations of systematic effects in the cold atom gravimeter,
rather than local variations of $g$.

\subsection{Sensitivity of the three gravimeters}

\begin{figure}[!t]
\centering
\includegraphics[width=4.2in,bb=0 0 996 357]{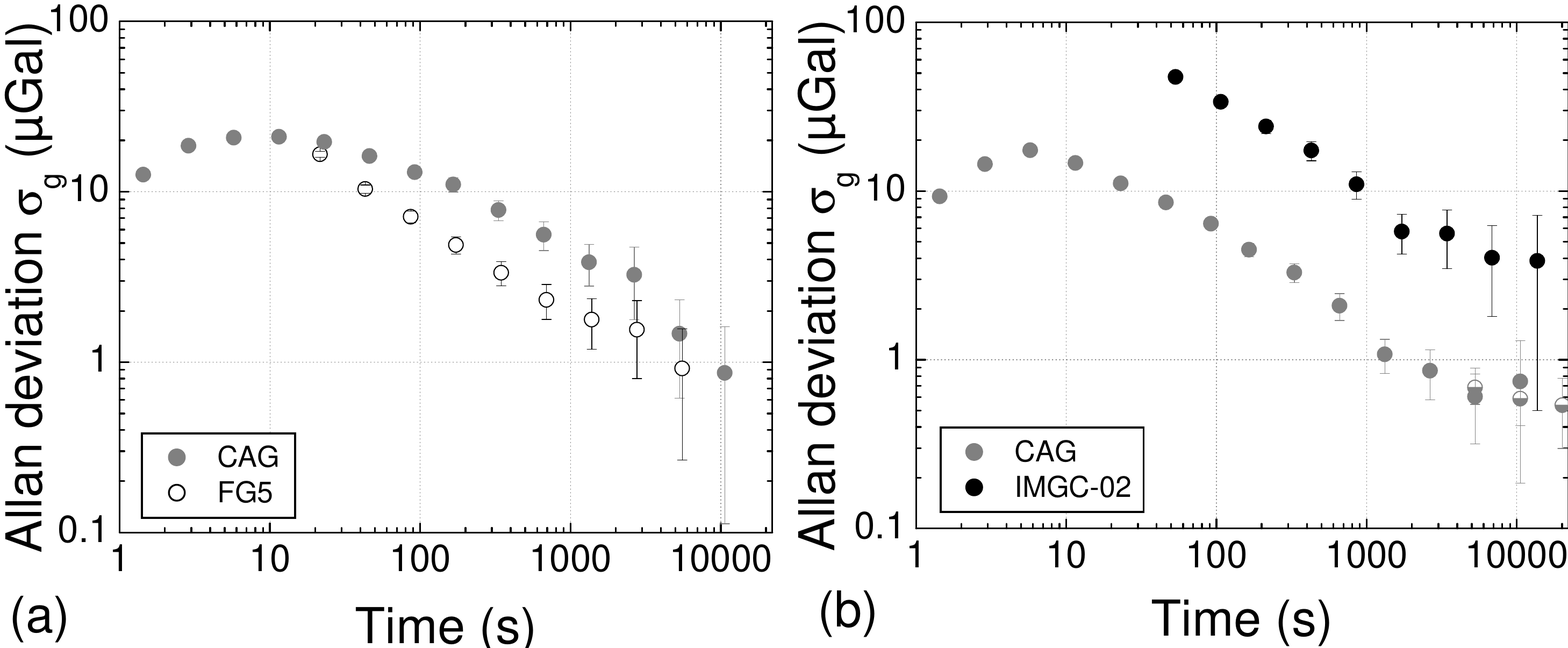}
\caption{Allan standard deviation of the corrected signals. (a)
FG5 and CAG on April 13th ($\textrm{MJD}=55299$); (b) CAG and
IMGC-02 on April 17th ($\textrm{MJD}=55303$). In graph (b) we also
plot (with semi-open circles) points corresponding to the Allan
deviation for the CAG for a longer duration, which indicate that
the CAG reaches a flicker floor around $(0.6\pm 0.3)~\mu$Gal after
$10000$~s.} \label{fig:allan}
\end{figure}

The sensitivity of a gravimeter is characterized by the Allan
standard deviation of the corrected $g$-measurement. In
Figure~\ref{fig:allan} we plot the Allan standard deviation for
the corrected $g$-measurements for the three absolute gravimeters.
Since we were not able to perform successful simultaneous $g$
measurements for all 3 gravimeters together, we calculate the
Allan deviations over the time periods indicated by the rectangles
in Figure~\ref{fig:goft}: FG5 and CAG around MJD$=55299$ (a), and
then CAG and IMGC-02 on MJD$=55303$ (b).

The Allan deviations of the three instruments scale as $t^{-1/2}$
(where $t$ is the cycle time of the measurements) which
corresponds to white noise. We compare the sensitivities of the
different gravimeters extrapolated to $1$s, following this white
noise behavior.

The optical interferometer IMGC-02 exhibits an equivalent
sensitivity at $1$s of $\sigma_g=330~\mu$Gal. For the FG5\#209,
$\sigma_g=70~\mu$Gal. The cold atom gravimeter exhibits a
sensitivity that is typically of $\sigma_g=140~\mu$Gal (as shown
in Figure~\ref{fig:allan}(a), on point GR40), but that can be as
good as $60~\mu$Gal (Figure~\ref{fig:allan}(b), on point GR8).
This variation in the noise level has already been mentioned in
paragraph~\ref{noises}.

\section{Conclusion}
In the context of the e-MASS project, we aim at determining the
gravitational acceleration $g$ with an accuracy at the $\mu$Gal
level, so that the contribution of the $g$ determination to the
watt balance uncertainty budget is negligible. The three
gravimeters that participated in the comparison reported here are
based on different methods, which is relevant for an accurate
determinaton of $g$.

This work is a significant step towards the achievement of the
gravimetry task defined in the e-MASS project. Indeed, in this
comparison, significant but repeatable differences between the 3
instruments have been measured, up to a level which is marginally
compatible with the claimed uncertainties. This indicates that
there are systematic effects that are not well evaluated yet. The
cold atom gravimeter still requires a more complete accuracy
budget determination, especially concerning the effect of optical
aberrations, which we take to be responsible for long-term
fluctuations of $g$ measurements. In particular, this aberration
shift fluctuates due to changes in the atomic trajectories, which
we plan to control better in the near future.

Additional comparisons with other absolute gravimeters based on
different methods will be organized, in order to progress towards
the absolute determination of $g$. The goal is to reach an
agreement at the $\mu$Gal level, in the perspective of completing
the gravimetric tasks of the e-MASS project.

\section*{Acknowledgments}
The authors would like to thank Franck Bielsa for fruitful discussions and for his invaluable help throughout the comparison.

The research within this EURAMET joint research project leading to these results has received funding from the European Community's Seventh Framework Programme, ERA-NET Plus, under Grant Agreement No. 217257, and from IFRAF (Institut Francilien pour les Atomes Froids).
Q. B. thanks CNES for supporting his work.

\ifCLASSOPTIONcaptionsoff
  \newpage
\fi

%
%

%
%
\newpage
%





\begin{thebibliography}{1}

\bibitem{geneves:emass}
    G. Genev\`es, F. Villar, F. Bielsa, O. Gilbert, A. Eichenberger, H. Baumann, G. D'Agostino, S. Merlet, F. Pereira Dos Santos, P. Pinot, P. Juncar, "The e-Mass Euramet Joint Research Project: the watt balance route towards a new definition of the kilogram", \emph{Digest of Conf. on Precision Electric Measurements (CPEM)}, Korea, 2010.
\bibitem{kibble}
    B. P. Kibble, "Atomic masses and fundamental constants", J. H. Sanders and A. H. Wapstra (Plenum, New York, 1976), vol. 5, pp. 545-551.
\bibitem{merlet2008}
    S. Merlet, A. Kopaev, M. Diament, G. Genev\`es, A. Landragin, F. Pereira Dos Santos, Micro-gravity investigations for the LNE watt balance project, \emph{Metrologia}, vol. 45, pp. 265-274, 2008.
\bibitem{baumann2009}
    H. Baumann, E. E. Klingel\'e, A. L. Eichenberger, P. Richard and B. Jeckelman, "Evaluation of the local value of the Earth gravity field in the context of the new definition of the kilogram", \emph{Metrologia} vol. 46, pp. 178-186, 2009.
\bibitem{merlet2007}
    S. Merlet, O. Francis, V. Palinkas, J. Kostelecky, N. Le Moigne, T. Jacobs, G. Genev\`es, "Absolute gravimetry measurements at LNE". \emph{Terrestrial Gravimetry Static and Mobile Measurements (TG-SMM 2007)}, Symp. Proc. (St Petersburg, Russia), pp. 173-174, 2007.
\bibitem{geneves:bw}
    G. Genev\`es, P. Gournay, A. Gosset, M. Lecollinet, F. Villar, P. Pinot, P. Juncar, A. Clairon, A. Landragin, D. Holleville, F. Pereira Dos Santos, J. David, M. Besbes, F. Alves, L. Chassagne and S. Topsu, "The BNM Watt Balance Project", \emph{IEEE Trans.  Instrum. and Meas.} vol. 54, pp. 850-853,  2005.
\bibitem{niebauer:fg5}
    T. M. Niebauer, G. S. Sasagawa, J. E. Faller, R. Hilt, F. Klopping, \emph{Metrologia}, vol. 32, pp. 159-180, 1995.
\bibitem{nelson:superspring}
    P. G. Nelson, "An active vibration isolation system for inertial reference and precision measurements", \emph{Rev. Sci. Instrum.} vol. 62, pp. 2069-2075, 1991.
\bibitem{eichenberger2009}
    A. Eichenberger, G. Genev\`es, P. Gournay, "Determination of the Planck constant by means of a watt balance", \emph{Eur. Phys. J. Special Topics.} 172, pp. 363-383, 2009.
\bibitem{dagostino2005}G. D'Agostino A. Germak, S. Desogus, C. Origlia, G. Barbato, "A method to estimate the time-position coordinates of a free-falling test-mass in absolute gravimetry", \emph{Metrologia}, vol. 42, pp. 233-238, 2005.
\bibitem{dagostino2008}
    G. D'Agostino A. Germak, S. Desogus, C. Origlia, G. Barbato, "Reconstruction of the free-falling body trajectory in a rise-and-fall absolute ballistic gravimeter", \emph{Metrologia} vol. 45, pp. 308-312, 2008.
\bibitem{bordé1989} Ch. J. Bord\'e, "Atomic interferometry with internal state labelling", Phys. Lett. A, vol.140, pp. 10-12, 1989.
\bibitem{legouet2008}
    J. Le Gou\"et, T. E. Mehlst\"aubler, J. Kim, S. Merlet, A. Clairon, A. Landragin, F. Pereira Dos Santos, "Limits to the sensitivity of a low noise compact atomic gravimeter", \emph{Appl. Phys. B}, 92, pp. 133-44, 2008.
\bibitem{dagostinoPhD}
    G. D'Agostino, "Development and metrological characterization of a new transportable absolute gravimeter", PhD thesis, Politecnico di Torino, Scuola di Dottorato, 2005.
\bibitem{fils2005}
    J. Fils, F. Leduc, Ph. Bouyer,  D. Holleville, N. Dimarcq, A. Clairon, A. Landragin, "Influence of optical aberrations in an atomic gyroscope", \emph{Eur. Phys. J. D} vol. 36, pp. 257-260, 2005.

\bibitem{gauguet2008} A. Gauguet, T. E. Mehlst\"aubler, T. L\'ev\`eque, J. Le Gou\"et, W. Chaibi, B. Canuel, A. Clairon, F. Pereira Dos Santos, A. Landragin. "Off-resonant Raman transition impact in an atom interferometer", \emph{Phys. Rev. A} vol. 78, pp. 043615, 2008.
\bibitem{merlet2010} S. Merlet, Q. Bodart, N. Malossi, A. Landragin, F. Pereira Dos Santos, O. Gitlein, L. Timmen, "Comparison between two mobile absolute gravimeters: optical versus atomic gravimeters",  \emph{Metrologia} vol. 47, ppL9-L11, 2010.
\end{thebibliography}
\end{document}